\documentclass[%
 aps,
 reprint,
 superscriptaddress,
 amsmath,amssymb,
]{revtex4-1}

\usepackage{graphicx}
\usepackage{dcolumn}
\usepackage{bm}
\usepackage{float}

\begin{document}

\title{Exceptional coupling in extreme skin-depth waveguides\\
for extremely low waveguide crosstalk}

\author{Md Borhan Mia}
\affiliation{Department of Electrical and Computer Engineering, Texas Tech University, Lubbock, Texas 79409, USA}

\author{Syed Z. Ahmed}
\affiliation{Department of Electrical and Computer Engineering, Texas Tech University, Lubbock, Texas 79409, USA}

\author{Ishtiaque Ahmed}
\affiliation{Department of Physics and Astronomy, Texas Tech University, Lubbock, Texas 79409, USA}

\author{Yun Jo Lee}
\affiliation{School of Electrical and Computer Engineering, Purdue University, West Lafayette, IN 47907, USA}

\author{Minghao Qi}
\affiliation{School of Electrical and Computer Engineering, Purdue University, West Lafayette, IN 47907, USA}

\author{Sangsik Kim}
\email{sangsik.kim@ttu.edu}
\affiliation{Department of Electrical and Computer Engineering, Texas Tech University, Lubbock, Texas 79409, USA}
\affiliation{Department of Physics and Astronomy, Texas Tech University, Lubbock, Texas 79409, USA}
\maketitle

\textbf{
Photonic chips can miniaturize complicate optical systems very tiny and portable, providing versatile functionalities for many optical applications
\cite{marin2017microresonator,atabaki2018integrating,dutt2018chip,sun2013large,miller2020large,kim2018photonic,hummon2018photonic,mehta2016integrated,spencer2018optical,kim2017dispersion,cheben2018subwavelength,li2017controlling,wang2019chip,halir2015waveguide}.
Increasing the photonic chip integration density is highly desired as it provides more functionalities, low cost, and lower power consumption
\cite{oulton2008hybrid,kim2015mode,haffner2015all,kim2014polarization,song2015high,gatdula2019guiding,shen2015metamaterial,shen2016increasing,gabrielli2012chip,Jahani2018ControllingIntegration}.
However, photonic chip integration density is limited by the waveguide crosstalk, which is caused by the evanescent waves in the cladding. Here we show that the waveguide crosstalk can be suppressed completely with the exceptional coupling in extreme skin-depth (eskid) waveguides. The anisotropic dielectric perturbations in the coupled eskid waveguides cause such an exceptional coupling, resulting in infinitely long coupling length. We demonstrate the extreme suppression of waveguide crosstalk via exceptional coupling on a silicon-on-insulator (SOI) platform, which is compatible with a complementary metal–oxide–semiconductor (CMOS) process. The idea of exceptional coupling in eskid waveguides
can be applied to many other photonic devices as well, significantly reducing entire chip footprints.
}

The waveguide crosstalk is the power transfer of a light signal between the adjacent waveguides due to the evanescent waves. In a typical photonic chip, to avoid the crosstalk, waveguides need to be separated large enough and this limits the integration density of photonic chips. To overcome this limit, plasmonics have been explored with their ability to confine light in subwavelength scale \cite{oulton2008hybrid,kim2015mode,kim2014polarization,haffner2015all}; however, there exist metallic losses to consider. Topological approaches that use a waveguide super-lattice \cite{song2015high,gatdula2019guiding}, inverse design \cite{shen2015metamaterial,shen2016increasing}, or transformations optics \cite{gabrielli2012chip} have been proposed,
yet these approaches add more complexity or phase variations in the design, and often lead to higher scattering losses. Recently, an extreme skin-depth (e-skid) waveguide that utilizes all-dielectric metamaterial claddings has been proposed to reduce the crosstalk \cite{Jahani2018ControllingIntegration}. The subwavelength multilayers effectively work as an anisotropic metamaterial and suppress the evanescent waves in the cladding. This reduces the crosstalk and an approximately 30 times longer coupling length has been demonstrated compared to typical strip waveguides. However, even with the reduced skin-depth, there still is some degree of crosstalk, and the further question remains as to if it is possible to suppress the crosstalk completely.

\begin{figure*}[!t]
\centering
\includegraphics[width=0.80\textwidth]{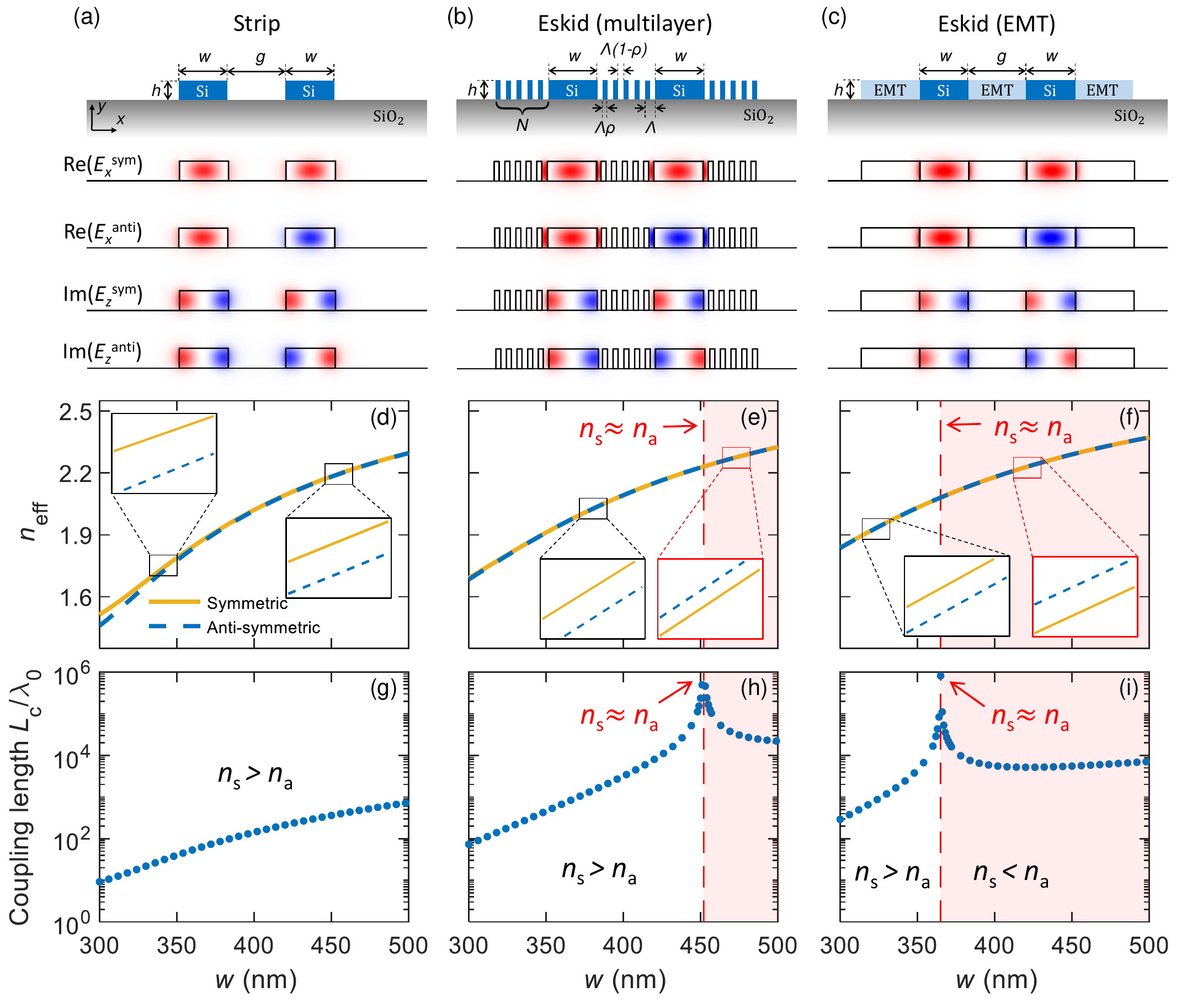}
\caption{
\textbf{On-chip coupled waveguide configurations and exceptional coupling in coupled extreme skin-depth (eskid) waveguides.} (a-c) Schematic cross-sections, geometric parameters, and mode profiles of the coupled silicon waveguides: (a) strip, (b) practical eskid with subwavelength multilayers, and (c) ideal eskid with effective medium theory (EMT). (d-f) Numerically simulated effective indices of the symmetric $n_{\rm s}$ (yellow solid) and anti-symmetric $n_{\rm a}$ (blue dashed) modes, and (g-i) their corresponding normalized coupling lengths $L_{\rm c}/\lambda_0=1/(2|n_{\rm s}-n_{\rm a}|)$ (blue dots): (d,g) strip, (e,h) eskid with multilayers, and (f,i) eskid with EMT. All the simulations are performed as a function of the core width $w$, while fixing the other parameters as $h=220$~nm, $\Lambda=100$~nm, $\rho=0.5$, and $N=5$. The free space wavelength is $\lambda_0=1550$~nm. The inset boxes of (d-f) show the zoomed-in view of each mode, and the red-shaded areas in (e,h) and (f,i) show the non-trivial coupling regimes, where $n_{\rm s}<n_{\rm a}$, which cannot be observed in a typical strip  coupling (d,g). The red arrows in (e,h) and (f,i) indicate the exceptional coupling points, where $n_{\rm s}\approx n_{\rm a}$, thus causing the $L_{\rm c}\rightarrow\infty$.
}
\label{fig:lc_neff}
\end{figure*}
To push the limit of chip integration density, we explore the exceptional coupling phenomena in the eskid waveguides by evaluating coupling lengths of the three different coupled waveguide schemes. Figures~\ref{fig:lc_neff}(a-c) show the cross-sections and geometric parameters of the three coupled waveguide configurations: (a) typical strip waveguides, (b) practical eskid waveguides with subwavelength-scale multilayer claddings, and (c) ideal eskid waveguides with anisotropic metamaterial claddings using the effective medium theory (EMT). All configurations are implemented on an SOI platform, i.e., Si and SiO$_2$ as a core and a substrate, respectively. Throughout the paper, we explore the coupled modes between the two identical fundamental quasi-transverse-electric (quasi-TE$_0$) modes, and the electric field profiles $\operatorname{Re}(E_x)$ and $\operatorname{Im}(E_z)$ of the coupled symmetric (sym) and anti-symmetric (anti) modes are also plotted in Figs.\ref{fig:lc_neff}(a-c). The multilayer eskid in Fig.~\ref{fig:lc_neff}(b) is a practical structure that can be fabricated with the current electron-beam lithography \cite{Jahani2018ControllingIntegration} and CMOS technology \cite{orcutt2011nanophotonic,stojanovic2018monolithic}. The EMT eskid in Fig.~\ref{fig:lc_neff}(c) is an equivalent model with an anisotropic metamaterial, and its permittivities ($\varepsilon_x=\varepsilon_\perp$ and $\varepsilon_y=\varepsilon_z=\varepsilon_\parallel$) follow \cite{milton2002theory,jahani2014transparent,Jahani2018ControllingIntegration},
\begin{subequations}
\label{eq:emt}
\begin{align}
    \varepsilon_{\parallel}&=\rho\varepsilon_{\rm Si}+(1-\rho)\varepsilon_{\rm air}\\
    \varepsilon_{\perp}&=\frac{\varepsilon_{\rm Si}\varepsilon_{\rm air}}
    {\rho\varepsilon_{\rm air}+(1-\rho)\varepsilon_{\rm Si}}
\end{align}
\end{subequations}
where, $\varepsilon_{\rm Si}$ and $\varepsilon_{\rm air}$ are the permittivities of Si and air, respectively. $\rho$ is the filling fraction of Si. Note that, due to the large index contrast between Si and air, a huge anisotropy can appear and its anisotropy can be engineered by controlling the $\rho$. With the increased anisotropy, the skin-depth in the cladding can be reduced, lowering the crosstalk \cite{jahani2014transparent,Jahani2018ControllingIntegration}.

The crosstalk between the two adjacent waveguides is assessed by the coupling length $L_{\rm c}$, which quantifies the length that transfers the optical power completely from one waveguide to the other waveguide; i.e., the crosstalk is lower for a longer $L_{\rm c}$, and it's the opposite for a shorter $L_{\rm c}$. To compare the coupling lengths of each configuration, effective refractive indices of the coupled symmetric ($n_{\rm s}$, yellow solid) and anti-symmetric ($n_{\rm a}$, blue dashed) modes are simulated in Figs.~\ref{fig:lc_neff}(d-f), and their corresponding normalized coupling lengths (blue dots) are plotted in Figs.~\ref{fig:lc_neff}(g-i): (d,g) strip, (e,h) eskid with multilayer, and (f,i) eskid with EMT. Each coupling length is normalized by the free-space wavelength at $\lambda_0=1550$~nm and the $L_{\rm c}$ of the two identical waveguides can be calculated by \cite{yariv2006photonics,huang1994coupled}
\begin{equation}
    \frac{L_{\rm c}}{\lambda_0}= \frac{1}{2\Delta n} = \frac{1}{2|n_{\rm s}-n_{\rm a}|}
    \label{eq:lc_neff}
\end{equation}
where, $\Delta n=|n_{\rm s}-n_{\rm a}|$ is the magnitude of the index difference between $n_{\rm s}$ and $n_{\rm a}$. The inset boxes in Figs.~\ref{fig:lc_neff}(d-f) show the zoomed-in view of $n_{\rm s}$ and $n_{\rm a}$ at different regimes. Notice that, in the coupled eskids of Figs.~\ref{fig:lc_neff}(e,f), there are non-trivial coupling regimes where $n_{\rm s}<n_{\rm a}$ (red-shaded region). This non-trivial coupling is not observable in typical strip waveguides (Fig.~\ref{fig:lc_neff}(d)). More importantly, at the transition from a typical coupling regime ($n_{\rm s}>n_{\rm a}$) to the non-trivial coupling regime ($n_{\rm s}<n_{\rm a}$), there is an exceptional coupling where $\Delta n$ approaches zero (i.e., $n_{\rm s}\approx{n_{\rm a}}$).
As shown in Figs.~\ref{fig:lc_neff}(h,i), at these exceptional coupling points, the coupling length approaches to infinity, i.e., the crosstalk is suppressed completely. For multilayer and EMT cases, the exceptional couplings appear at different $w$. This is due to the deviations of effective $\varepsilon_{\perp}$ and $\varepsilon_{\parallel}$ of the multilayer compared to those of ideal EMT (see Supplementary).

\begin{figure*}[!ht]
\centering
\includegraphics[width=0.80\textwidth]{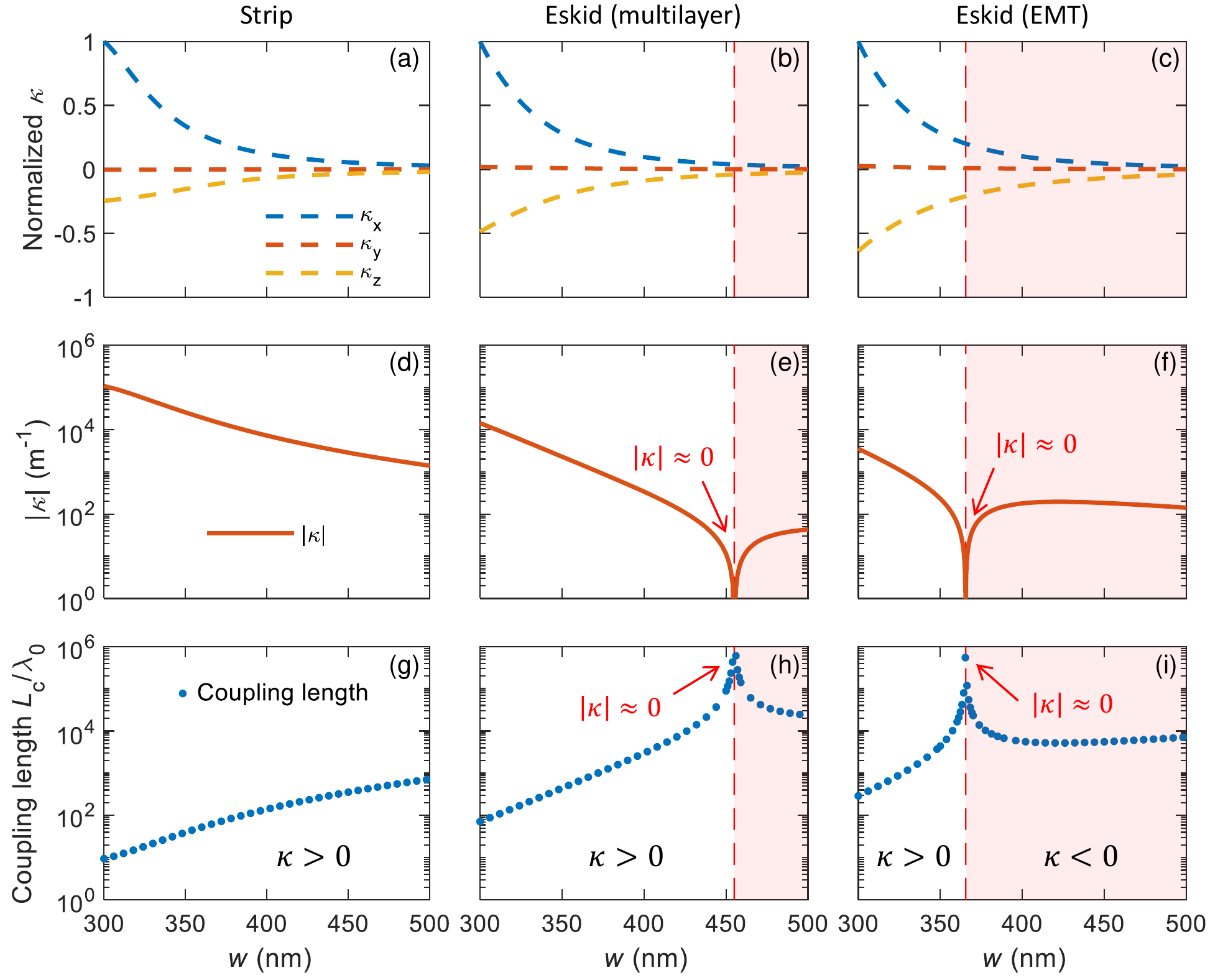}
\caption{
\textbf{Anisotropic coupled mode analysis on the exceptional coupling in coupled eskid waveguides.} (a-c) Normalized anisotropic coupling coefficients $\kappa_x$ (blue dashed), $\kappa_y$ (orange dashed), and $\kappa_z$ (yellow dashed) of the coupled (a) strip, (b) eskid with multilayer, and (c) eskid with EMT waveguides. Geometric parameters and the wavelength are the same as in Figs.~\ref{fig:lc_neff}. (d-f) Magnitude of the total coupling coefficient $|\kappa| = |\kappa_x + \kappa_y+ \kappa_z|$ (orange solid), and (g-i) their corresponding normalized coupling lengths $L_{\rm c}/\lambda_0=\pi/(2|\kappa|\lambda_0)$ (blue dots) for each configuration: (d,g) strip, (e,h) eskid with multilayer, and (f,i) eskid with EMT. The normalized coupling lengths (g-i) that are obtained with anisotropic coupled mode analysis match with those results in Figs.~\ref{fig:lc_neff}(g-i) from the full numerical simulations. The red-shaded areas in (e,h) and (f,i) show the non-trivial coupling regimes where $\kappa<0$, which cannot be observed in typical strip waveguide coupling (d,g). The red arrows in (e,h) and (f,i) indicate the exceptional coupling points where $|\kappa|\approx0$, thus causing the $L_{\rm c}\rightarrow\infty$. As shown in (b) and (c), the anisotropic nature of eskid waveguides can cause a larger $\kappa_z$, which results in the non-trivial coupling regime ($\kappa<0$) and the exceptional coupling ($\kappa\approx 0$) at the transition.
}
\label{fig:lc_kappa}
\end{figure*}
To understand the underlying mechanism of the exceptional coupling, we analyzed each configuration using the anisotropic coupled mode analysis. In a quasi-TE$_0$, an $E_x$ component is dominant but there is an $E_z$ component as well. Thus, to address the coupled modes correctly, the anisotropic coupling coefficients from all the field components (i.e., $\kappa_x, \kappa_y$, and $\kappa_z$) should be considered(see Methods). The overall coupling coefficient $|\kappa|$ can be obtained by adding each component together (i.e., $|\kappa|=|\kappa_x+\kappa_y+\kappa_z|$) and the coupling length of the two same waveguides is the following \cite{yariv2006photonics,huang1994coupled}:
\begin{equation}
    L_c=\frac{\pi}{2|\kappa|}
\label{eq:lc_kappa}
\end{equation}
Figures~\ref{fig:lc_kappa}(a-c) show the normalized coupling coefficients of each component $\kappa_x$ (blue dashed), $\kappa_y$ (orange dashed), and $\kappa_z$ (yellow dashed), and their corresponding overall coupling coefficient $|\kappa|$ (orange solid) and the normalized coupling length $L_{\rm c}/\lambda_0$ (blue dots) are plotted in Figs.~\ref{fig:lc_kappa}(d-f) and Figs.~\ref{fig:lc_kappa}(g-i), respectively: (a,d,g) coupled strip, and coupled eskid with (b,e,h) the multilayer and (c,f,i) EMT. In every case, as the $w$ increases, the coupling coefficients are reduced and the coupling lengths are increased, due to the higher confinement in the core and less overlap between the modes. In Fig.~\ref{fig:lc_kappa}(a), the $\kappa_x$ is clearly dominant than the other components, even with a non-negligible $\kappa_z$. The sign of $\kappa_z$ is negative due to the imaginary $E_z$, and it counteracts with the $\kappa_x$ in determining the $|\kappa|$. In coupled strip waveguides, the magnitude of $\kappa_x$ is always greater than that of $\kappa_z$ (i.e., $\kappa>0$) as the $E_x$ is dominant in the quasi-TE$_0$ mode. Figure~\ref{fig:lc_kappa}(d) shows the overall $|\kappa|$ with the actual unit, and its corresponding normalized coupling length in Fig.~\ref{fig:lc_kappa}(g) exactly matches the result from the full numerical simulation in Fig.~\ref{fig:lc_neff}(g).
\begin{figure*}[!t]
\centering
\includegraphics[width=1.00\textwidth]{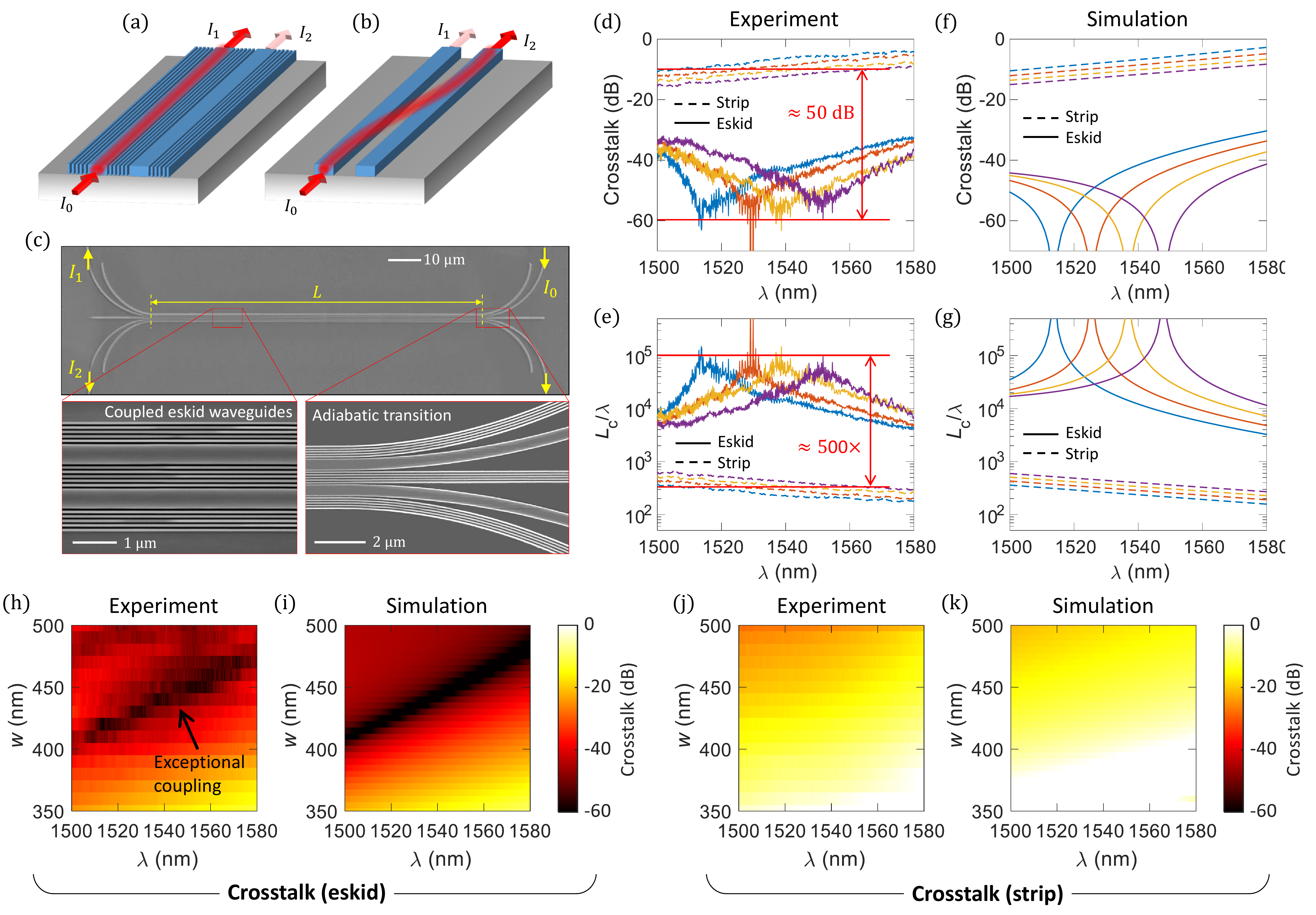}
\caption{
\textbf{Experimental demonstration of the exceptional coupling in coupled eskid waveguides.}
Schematic view of the coupled (a) eskid (multilayer) and (b) strip waveguides. $I_0, I_1$, and $I_2$ indicate the optical powers at input, through, and coupled ports, respectively. (c) SEM images of the fabricated devices. Zoomed-in images show (left) the coupled eskid waveguides and (right) the adiabatic transition from strip to eskid waveguides. (d) Experimentally measured waveguide crosstalk and (e) the corresponding normalized coupling length of the coupled eskid (solid) and strip (dashed) waveguides: $w=420$~nm (blue), 430~nm (orange), 440~nm (yellow), and 450~nm (purple). Numerically simulated (f) crosstalk and (g) normalized coupling length that correspond to the experimental results in (d) and (e), respectively. Geometric parameters are $h=220$~nm, $\rho=0.5$, $\Lambda=100$~nm, $N=5$, and $L=100~\mu$m. (h-k) Map plots of the crosstalk as functions of $\lambda$ and $w$ for the coupled (h,i) eskid and (j,k) strip waveguides: (h,j) Experiment and (i,k) Simulation. Dark regions in (h,i) indicate the exceptional couplings in coupled eskid waveguides.
}
\label{fig:exp}
\end{figure*}
In cases of eskids, there are non-trivial coupling regimes, where the magnitude of $\kappa_z$ is greater than that of $\kappa_x$ (i.e., $\kappa<0$) as shown in Figs.~\ref{fig:lc_kappa}(b) and \ref{fig:lc_kappa}(c) (red-shaded). These non-trivial coupling regimes in the coupled eskids are due to the anisotropic dielectric perturbations $\Delta\varepsilon_i$ of the scheme
(i.e., $\Delta\varepsilon_x\neq\Delta\varepsilon_y=\Delta \varepsilon_z$), allowing the $\kappa_z$ to compensate for the $\kappa_x$. The overall $|\kappa|$ approaches to zero at the transition points, resulting in infinitely long coupling lengths (i.e., $L_{\rm c}\rightarrow\infty$) as in Figs.~\ref{fig:lc_kappa}(h) and \ref{fig:lc_kappa}(i); these results also match well with the full simulation results in Figs.~\ref{fig:lc_neff}(h) and \ref{fig:lc_neff}(i), respectively.

To confirm our theoretical finding, we fabricated the coupled eskid (multilayer) and strip waveguides, and the crosstalks of each configuration are measured and compared. Figures~\ref{fig:exp}(a) and \ref{fig:exp}(b) show the schematic views of the coupled eskid and strip waveguides, respectively, and Fig.~\ref{fig:exp}(c) shows the SEM images of the fabricated devices. The $I_0$ indicates the optical power at the input port, and the $I_1$ and $I_2$ are the output powers at the through and coupled ports, respectively. The crosstalk is defined as the power ratio $I_2/I_1$ and it is related to the coupling length $L_c$, following \cite{yariv2006photonics},
\begin{equation}
\frac{I_2}{I_1}=\tan^2\left(\frac{\pi L}{2L_c}\right)
\label{eq:crosstalk}
\end{equation}
where $L$ is the physical length of the coupled waveguides. To measure the crosstalk, we sent a light signal to the input port $I_0$ through a fiber-coupled grating coupler, then output signals $I_1$ and $I_2$ were measured simultaneously (see Methods). Figure~\ref{fig:exp}(d) shows the experimentally measured crosstalk (in dB) and Fig.~\ref{fig:exp}(e) is the corresponding normalized coupling length. Solid and dashed lines are the cases of the coupled eskid and strip waveguides, respectively, and each color represents different core widths $w$.
Figures~\ref{fig:exp}(f) and \ref{fig:exp}(g) are the simulation results that correspond to Figs.~\ref{fig:exp}(d) and \ref{fig:exp}(e), respectively. Note that the dips in the crosstalk and the peaks in the $L_c/\lambda$ indicate the exceptional couplings. In Fig.~\ref{fig:exp}(d), notice that the crosstalks of the coupled eskids are suppressed down to about $-60$~dB, which is approximately 50~dB lower than that of the standard strip waveguides. As to the coupling length, the peak $L_c/\lambda$ of the coupled eskids are in the order of $10^5$, which is approximately $500$ times longer than the case of strip waveguides. As shown in Figs.~\ref{fig:exp}(f) and \ref{fig:exp}(g), in an ideal case of exceptional coupling, the crosstalk can be suppressed completely with an infinitely long coupling length. However, in real experiments, the minimum crosstalk is limited by the scattering from the waveguide sidewall roughness and the cross-coupling at the transition between strip to eskid waveguides(see Supplementary). Still, the crosstalk that we achieved here is extremely low, and, to the best of our knowledge, these results demonstrate the longest coupling length, spanning about $10^5$ of free-space wavelengths.

The full map plots of the crosstalk, as functions of $\lambda$ and $w$, for the coupled eskid and strip waveguides are plotted in Figs.~\ref{fig:exp}(h,i) and \ref{fig:exp}(h,i), respectively, clearly showing much lower crosstalk with the eskid waveguides.  Figures~\ref{fig:exp}(h,j) and \ref{fig:exp}(i,k) are the experimental and simulation results, respectively. The dark regions in Figs.~\ref{fig:exp}(h,i) indicate the exceptional coupling, which can be observed only with the coupled eskid waveguides. Since the exceptional coupling occurs at the point where the coupling coefficient $\kappa_z$ compensates the $\kappa_x$, the exceptional coupling can be engineered by controlling the modal overlaps between the two waveguides. For example, as shown in Figs.~\ref{fig:exp}(h,i), increasing the $w$ shifts the exceptional coupling to a longer wavelength; this is because a wider $w$ increases the light confinement and reduces the modal overlap between the two coupled eskids, while a longer wavelength works the opposite way. Similarly, changing the other geometric parameters $h, g,$ and $\rho$ shifts the exceptional coupling point, and we also observed exceptional couplings
with different numbers of eskid layers $N$ and filling fraction $\rho$ (see Supplementary).

In summary, we have presented exceptional couplings in the coupled eskid waveguides that can achieve extremely low waveguide crosstalk. Our coupled mode analysis reveals that the unique anisotropic dielectric perturbation of the eskid is the fundamental origin of the non-trivial coupling regime that can cause the exceptional coupling at the transition. We experimentally demonstrated the exceptional couplings on an SOI platform, which is low-loss, low-cost, and compatible with the CMOS foundry. We suppressed the crosstalk approximately 50~dB lower than
the case of strip waveguides, which corresponds to an approximately 500 times longer coupling length. The exceptional coupling can be engineered with geometric parameters and the filling fraction of the eskids. Our approach of using the exceptional coupling in the coupled eskid waveguides drastically increase the photonic chip integration density and can be applied to other photonic devices realizing highly dense PICs.

\section*{Methods}
\noindent {\bf Numerical simulation.}
A commercially available software (Lumerical Mode Solution) was used to calculate the effective refractive indices of the coupled symmetric and anti-symmetric modes. For the implementation of an ideal metamaterial cladding, the EMT was used for the $\varepsilon_x=\varepsilon_{\perp}$ and $\varepsilon_y=\varepsilon_z=\varepsilon_{\parallel}$, following Eq.~(\ref{eq:emt}).
For the strip-to-eskid adiabatic transition, a full 3D FDTD simulation was used to minimize the cross-coupling efficiency (see Supplementary).\\

\noindent {\bf Coupled mode analysis.}
For the coupled mode analysis in Fig.~\ref{fig:lc_kappa}, we calculated the $\kappa_i$, following
\cite{yariv2006photonics,huang1994coupled}:
\begin{equation}
\kappa_i= \frac{\omega\varepsilon_0}{4}\iint \Delta\varepsilon_i(x,y)
            {E_{1i}}(x,y) {E_{2i}^{*}}(x,y) dxdy
\label{eq:kappa}
\end{equation}
where the subscript $i=x,y,$ and $z$. ${E_{1i}}$ and ${E_{2i}}$ are the normalized electric fields of isolated quasi-TE$_0$ modes at each side and $\Delta \varepsilon_i$ is the dielectric perturbation between them. Note that, for isotropic media, as in strip waveguides, all the dielectric perturbation components are the same (i.e., $\Delta\varepsilon_x=\Delta\varepsilon_y=\Delta\varepsilon_z$); however, for anisotropic cases, as in eskid waveguides, they are different (i.e., $\Delta\varepsilon_x\neq\Delta\varepsilon_y=\Delta \varepsilon_z$), causing the non-trivial coupling regime. Each coupling coefficient $\kappa_i$ was added together to form the total coupling coefficient $|\kappa|$, and Eq.~(\ref{eq:lc_kappa}) was used to calculate the corresponding coupling length.\\

\noindent {\bf Device fabrication.}
The photonic chips were fabricated on an SOI wafer ($220$~nm thick Si on a $2~\mu$m SiO$_2$) using the JEOL JBX-6300 EBL system, which operated at $100$~KeV energy, $400$~pA beam current,
and $500~\mu$m $\times$ $500~\mu$m exposure field. A solvent rinse was done, followed by $5$~min of O$_2$ plasma treatment. Hydrogen silsequioxane resist (HSQ, Dow-Corning XR-1541-006) was spin coated at $4000$~rpm and pre-exposure baked on a hotplate at $90^{\circ}$ for $5$~min. Shape placements by the machine grid, the beam stepping grid, and the spacing between dwell points during shot shape writing were $1$~nm, $4$~nm, and $4$~nm, respectively. An exposure dose of $1460$ $\rm {\mu C/cm^2}$ was used. The resist was developed in $25\%$ tetramethylammonium hydroxide (TMAH) for $4$~min followed by a flowing deionized waster rinse for $60$~s and an isopropanol rinse for $10$~s. Then, nitrogen was blown to air dry. After development of the resist, the unexposed top silicon layer was etched by a Cl$_2$/O$_2$ in a reactive ion-plasma etching tool (Trion Minilock) to transfer the pattern from the resist to the silicon layer.\\

\noindent {\bf Device characterization.}
The photonic chips were characterized by a custom-built grating coupler setup. An angle polished ($8^{\circ}$) eight-channel fiber array was used to couple light in and out of the grating couplers. The fiber array was mounted on a five-axis stage with a high-precision adjuster with 20~nm sensitivity in XYZ direction. A Keysight Tunable Laser 81608A was used as the source and a Keysight N7744A optical power meter with InGaAs sensors was used as the output detector. The wavelength was swept from 1500 to 1580~nm with a step of 100~pm. A polarization controller was used to control the polarization of the input laser light.

\begin{acknowledgments}
This material is based upon work supported by the National Science Foundation under Grant No. ECCS-1930784. This work was performed, in part, at the Center for Integrated Nanotechnologies, an Office of Science User Facility operated for the U.S. Department of Energy (DOE) Office of Science by Los Alamos National Laboratory (Contract 89233218CNA000001) and Sandia National Laboratories (Contract DE-NA-0003525).
\end{acknowledgments}

\section*{Author contributions}
S.K. conceived the project and guided the theoretical and experimental investigations. M.M. conducted all the numerical simulations and the coupled mode analysis. S.A. measured all the devices. I.A. fabricated all devices used in the reported experiments. Y.L. and M.Q. supported the project with initial sets of devices. M.M. and S.K. wrote the manuscript. All discussed and commented on the results.

\bibliographystyle{naturemag}
\nocite{*}

\end{document}


\title{Supplementary Information for ``Exceptional coupling in extreme skin-depth waveguides for extremely low waveguide crosstalk''}

\author{Md Borhan Mia}
\affiliation{Department of Electrical and Computer Engineering, Texas Tech University, Lubbock, Texas 79409, USA}

\author{Syed Z. Ahmed}
\affiliation{Department of Electrical and Computer Engineering, Texas Tech University, Lubbock, Texas 79409, USA}

\author{Ishtiaque Ahmed}
\affiliation{Department of Physics and Astronomy, Texas Tech University, Lubbock, Texas 79409, USA}

\author{Yun Jo Lee}
\affiliation{School of Electrical and Computer Engineering, Purdue University, West Lafayette, IN 47907, USA}

\author{Minghao Qi}
\affiliation{School of Electrical and Computer Engineering, Purdue University, West Lafayette, IN 47907, USA}

\author{Sangsik Kim}
\email{sangsik.kim@ttu.edu}
\affiliation{Department of Electrical and Computer Engineering, Texas Tech University, Lubbock, Texas 79409, USA}
\affiliation{Department of Physics and Astronomy, Texas Tech University, Lubbock, Texas 79409, USA}

\maketitle
\newpage
\section{Effective medium theory vs. subwavelength multilayers}
\begin{figure}[!ht]
\centering
\includegraphics[width=0.8\linewidth]{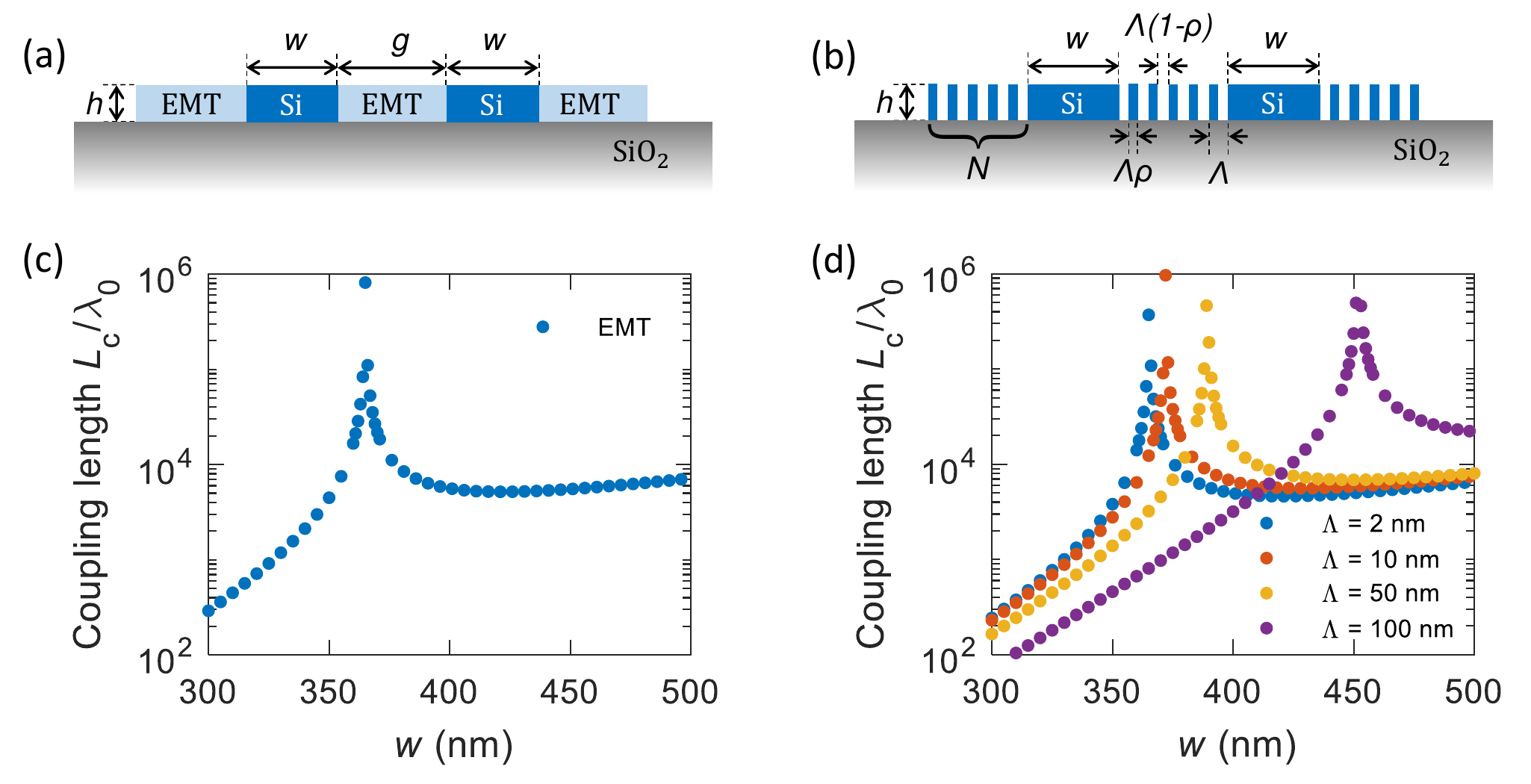}
\caption{{\bf Ideality of anisotropic metamaterial.}
Schematics of the (a) ideal eskid with effective medium theory (EMT) and (b) practical eskid with multilayer claddings. Normalized coupling lengths $L_c/\lambda_0$ for the coupled eskid waveguides with (c) EMT and (d) multilayers with different periodicity: $\Lambda=2$~nm (blue dots), $10$~nm (orange dots), $50$~nm (yellow dots), and $100$~nm (purple dots). The filling fraction is set to $\rho=0.5$ and the other parameters are $h=220$~nm, $g=550$~nm, and $\lambda_0=1550$~nm. Note that, as the $\Lambda$ reduces, the exceptional coupling point with multilayers in (d) approaches to that with EMT in (c). This is because the subwavelength-scale multilayer structure approaches an ideal EMT metamaterial as the $\Lambda$ reduces (i.e, $\Lambda<<\lambda$) \cite{milton2002theory}.
}
\label{fig:eskid_emt}
\end{figure}
\newpage
\section{Strip-to-eskid adiabatic mode transition}
\begin{figure}[!ht]
\centering
\includegraphics[width=0.95\linewidth]{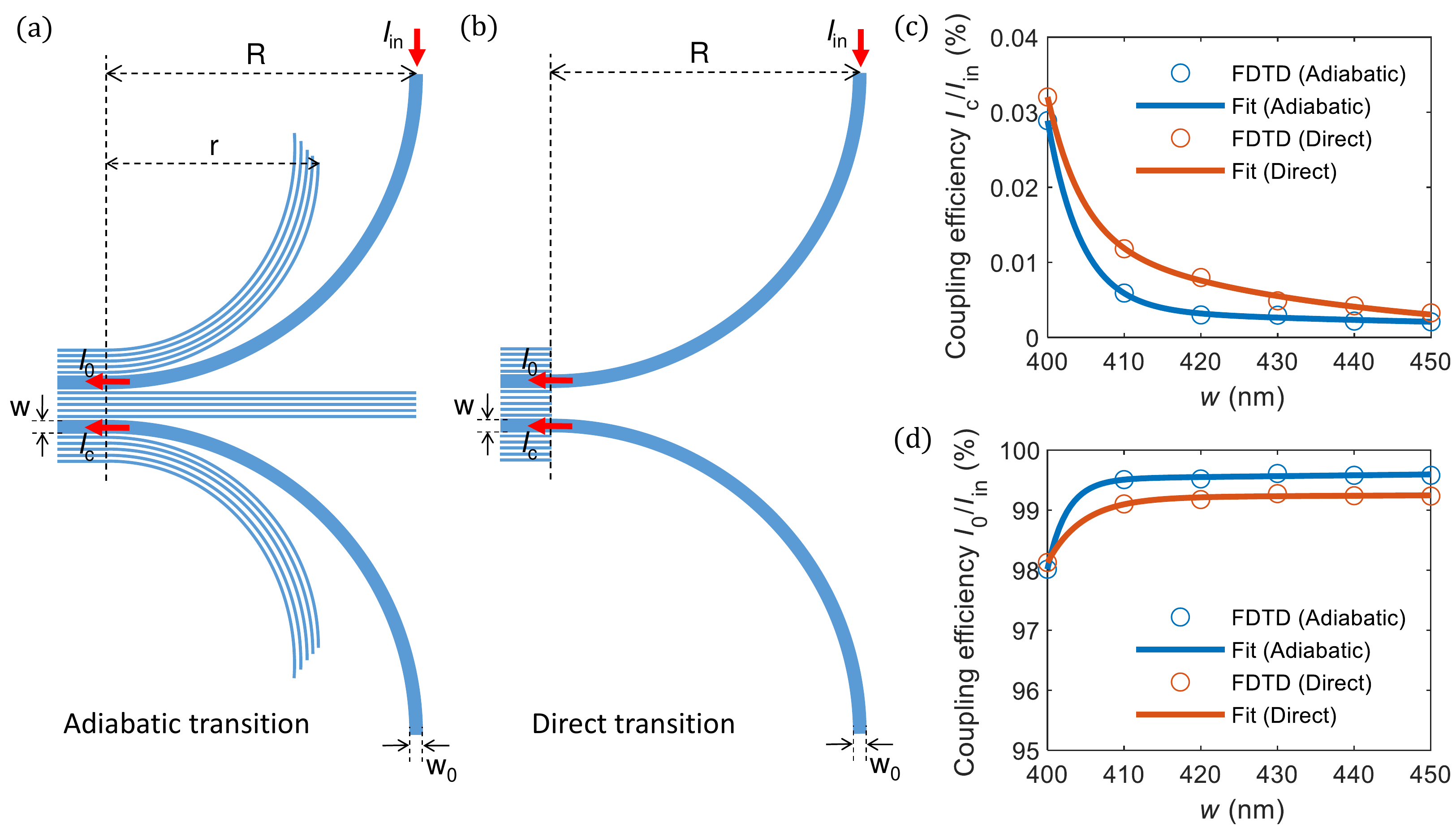}
\caption{\textbf{Strip-to-eskid interface optimization}. (a,b) Schematics of the strip-to-eskid mode transition region: (a) Adiabatic and (b) Direct transitions. The bending radii are $R=20~\mu$m and $r=15~\mu$m, respectively. The core width at the input strip waveguide is fixed to $w_0=450$~nm, while the core width of the eskid is varied $w=400-450$~nm. The core width is adiabatically tapered through the bending. $I_{\rm in}$ is the input power at the strip waveguide, whereas $I_{\rm 0}$ and $I_{\rm c}$ are the powers at the through and coupled ports of the eskid.(c,d) Simulated power coupling efficiencies of (c) $I_{c}/I_{\rm in}$ and (d) $I_{0}/I_{\rm in}$ as a function of $w$ for adiabatic (blue) and direct (orange) transitions. Circles indicate the full 3D FDTD simulation results and solid lines are their fitting curves. Notice that the adiabatic transition reduces the coupling efficiency $I_{c}/I_{\rm in}$ close to $10^{-5}$. This allowed us to measure the extremely low waveguide crosstalk in the experiment.
}
\label{fig:interface}
\end{figure}
To measure the waveguide crosstalk at the extremely low power level, the interface crosstalk at each port should be lower than the waveguide crosstalk. When measuring waveguide crosstalk above $-40$~dB, one may use a direct strip-to-eskid coupling as shown in Fig.~\ref{fig:interface}(b) \cite{Jahani2018ControllingIntegration,han2016strip}. However, in our case, the power level of the waveguide crosstalk due to the exceptional coupling is below $-40$~dB and an additional transition scheme is required. Figures~\ref{fig:interface}(a) and \ref{fig:interface}(b)
show the strip-to-eskid interface schemes with adiabatic and direct transitions, respectively. $R$ is the bending radius of the waveguide core and $r$ is the bending radius of the metamaterial multilayers. The powers at the input strip, through port eskid, and coupled eskid waveguides are denoted by $I_{\rm in}$, $I_{\rm 0}$, and $I_{\rm c}$, respectively. We fixed the core width at the input strip waveguide to $w_0=450$~nm and the core width of eskid waveguide $w$ is varied from $400$~nm to $450$~nm. Through the bending, the core width is tapered adiabatically. Figures~\ref{fig:interface}(c) and \ref{fig:interface}(d) are the full 3D FDTD simulation results showing the coupling efficiencies for $I_{c}/I_{\rm in}$ and $I_{0}/I_{\rm in}$, respectively, with the banding radii of $R=20~\mu$m and $r=15~\mu$m. Blue and orange circles are the coupling efficiencies of the adiabatic and direct transition schemes, respectively, and solid lines are their fitting curves. Since the exceptional couplings are observed at the power level below $-40$~dB, the coupling efficiency $I_{c}/I_{\rm in}$
should be less than $10^{-4}$. With the direct transition, the coupling efficiency $I_{c}/I_{\rm in}$ is at the border-line of $10^{-4}$ and the exceptional coupling phenomena were not seen clearly (i.e., too shallow dips). However, with the adiabatic transition as in Fig.~\ref{fig:interface}(c), the coupling efficiency $I_{c}/I_{\rm in}$ has been suppressed further down to  $\approx10^{-5}$ and the exceptional coupling phenomena were observed clearly as shown in Fig.~3 of the main manuscript.

\newpage
\section{Parametric studies on the exceptional coupling}
\vspace{-10pt}
\begin{figure}[!ht]
\centering
\includegraphics[width=0.9\linewidth]{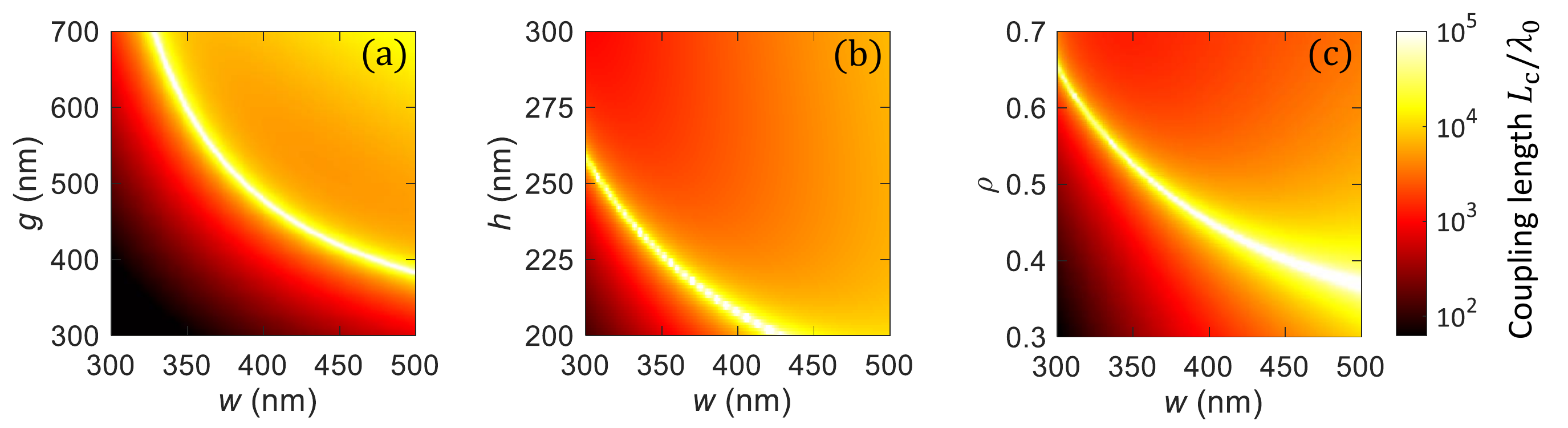}
\caption{
\textbf{Parametric studies on the exceptional coupling.} Normalized coupling lengths $L_\textrm{c}/\lambda_{0}$ of the coupled EMT eskid waveguides as functions of (a) $w$ and $g$,(b) $w$ and $h$, and (c) $w$ and $\rho$. Other parameters are set to $h=220$~nm, $g=550$~nm, and $\rho=0.5$, unless otherwise specified. The free-space wavelength is $\lambda_0=1550$~nm.
}
\vspace{-10pt}
\label{fig:parameter}
\end{figure}
As the exceptional coupling exists near the $|\kappa_x|\approx|\kappa_z|$, which is determined by the anisotropic dielectric perturbation and the modal overlap, changing the filling fraction $\rho$ and geometric parameters $w,g$, and $h$ would shift the exceptional coupling point. In other words, these parameters would work as tuning knobs to engineer the exceptional coupling points. To explore the engineering capability of the exceptional coupling, full parametric simulations were conducted on the coupled eskid configuration with EMT. Figures~\ref{fig:parameter}(a-c) show the calculated $L_{\rm c}/\lambda_0$ map plots as functions of (a) $w$ and $g$, (b) $w$ and $h$, and (c) $w$ and $\rho$, respectively. Other parameters are set to $h=220$~nm,  $\rho=0.5$, $g=550$~nm, and $\lambda_0=1550$~nm, unless otherwise specified. In Fig.~\ref{fig:parameter}(a), it is noted that, as the $g$ reduces, the exceptional coupling appears at a wider $w$. This is due to the similar trend in $g$ and $w$ for the modal overlap; as the $g$ reduces, there will be more modal overlap due to the proximity, while there will be less modal overlap as the $w$ increases due to the higher light confinement. The reduced modal overlap due to the increased $w$ compensates for the increased modal overlap due to the reduced $g$, shifting the exceptional coupling point. Changing the $h$ of the scheme would show a similar effect as the $g$ and $w$, as increasing the $h$ allows for a higher confinement, thus less modal overlap. To compensate for the reduced modal overlap due to the increased $h$, the $w$ should be narrower. This trend is clearly shown in Fig.~\ref{fig:parameter}(b). Changing the $\rho$ is more complicate than changing other parameters, as it simultaneously modifies both $\varepsilon_\perp$ and $\varepsilon_\parallel$. However, within the range of our evaluation, the anisotropy increases as $\rho$ increases and the effect of anisotropic dielectric perturbation becomes more dominant. Thus, the $\kappa_z$ with a higher $\rho$ can compensate for the $\kappa_x$ at a larger modal overlap, which is a narrower $w$. This trend is also clearly shown in Fig.~\ref{fig:parameter}(c).

\newpage
\section{Exceptional couplings with different parameters}
\begin{figure*}[!ht]
\centering
\includegraphics[width=1.0\textwidth]{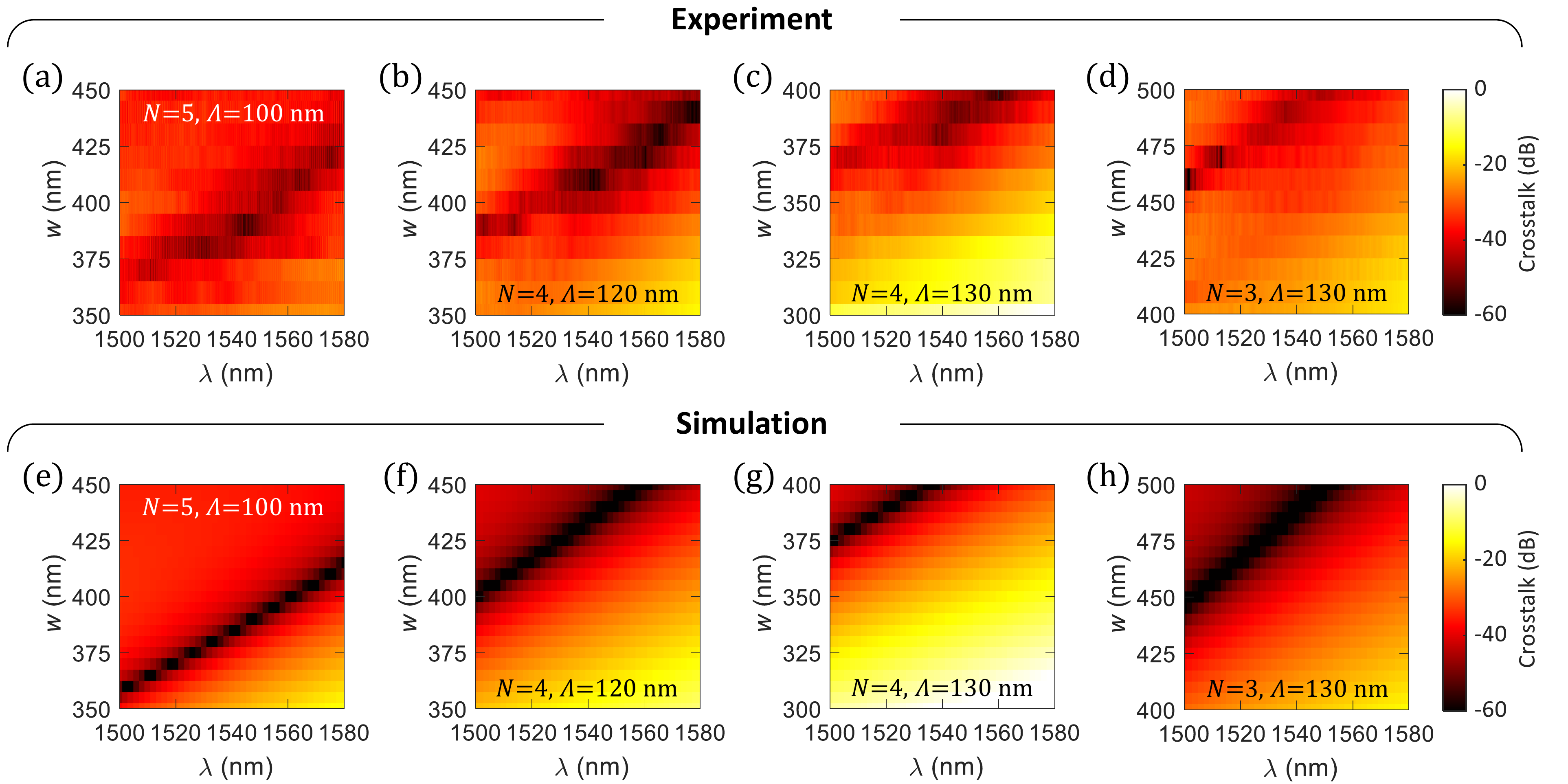}
\caption{\textbf{Exceptional couplings with different geometric parameters}. (a-d) Experimentally measured waveguide crosstalk maps as functions of $\lambda$ and $w$ for different geometric parameters: (a) $N=5$, $\Lambda=100$~nm, $\Lambda(1-\rho)=40$~nm, (b) $N=4$, $\Lambda=120$~nm, $\Lambda(1-\rho)=50$~nm, (c) $N=4$, $\Lambda=130$~nm, $\Lambda(1-\rho)=50$~nm, and (d) $N=3$, $\Lambda=130$~nm, $\Lambda(1-\rho)=50$~nm. The height is fixed to $h=220$~nm. (e-h) Simulated crosstalk maps that correspond to (a-d).
}
\label{fig:diffP}
\end{figure*}
As shown in Fig.~\ref{fig:parameter}, the exceptional coupling can be engineered with geometric parameters that determine the modal overlap and the anisotropic dielectric perturbation. In cases of the practical eskid waveguides with subwavelength-scale multilayers, the number of eskid layers $N$ and the periodicity $\Lambda$ determine the gap $g$ between the two eskid waveguides, and the ratio of the multilayer width $\Lambda\rho$ and gap $\Lambda(1-\rho)$ define the filling fraction $\rho$. For the practical implementation of the eskid waveguide, the minimum feature size limits the gap size of multilayers $\Lambda(1-\rho)$ to be larger than 40~nm. Considering these limitations, we also demonstrated exceptional couplings with different sets of $N$ and $\rho$. Figures~\ref{fig:diffP}(a-d) show the crosstalk map plots as functions of $\lambda$ and $w$ for different geometries: (a) $N=5$, $\Lambda=100$~nm, $\Lambda(1-\rho)=40$~nm, (b) $N=4$, $\Lambda=120$~nm, $\Lambda(1-\rho)=50$~nm, (c) $N=4$, $\Lambda=130$~nm, $\Lambda(1-\rho)=50$~nm, and (d) $N=3$, $\Lambda=130$~nm, $\Lambda(1-\rho)=50$~nm. Note that the filling fractions of each case are (a) $\rho=0.6$, (b) $\rho=0.583$, (c) $\rho=0.615$, and (d) $\rho=0.615$. Figures~\ref{fig:diffP}(e-h) are the simulation results that correspond to Figs.~\ref{fig:diffP}(a-d). Notice that, in Fig.~\ref{fig:diffP}(a) ($\rho=0.6$), the exceptional coupling appears at a narrower $w$ compared to the case of Fig.~3(h) ($\rho=0.5$) in the main manuscript. The increased $\rho$ introduces a higher anisotropic dielectric perturbation, allowing the $\kappa_z$ to compensate for $\kappa_x$ at a larger modal overlap,
i.e., a narrower $w$. This trend is consistent with our parametric analysis in Fig.~\ref{fig:parameter}(c). A similar trend is shown between Figs.~\ref{fig:diffP}(b) and \ref{fig:diffP}(c);
the $\rho=0.615$ of Fig.~\ref{fig:diffP}(c) is higher than $\rho=0.583$ of Fig.~\ref{fig:diffP}(b), shifting the exceptional coupling point to a narrower $w$. The $g$ of Fig.~\ref{fig:diffP}(c) is also slightly larger ($\Delta g=40$~nm) than that of Fig.~\ref{fig:diffP}(b), having the same effect, i.e., a narrower $w$. To separately observe the effect of $g$,
we also tested the devices with the $N=3$ in Fig.~\ref{fig:diffP}(d). Note that, between Figs.~\ref{fig:diffP}(c) and \ref{fig:diffP}(d), the only difference is $N$ while the other parameters are the same. Thus, we can view the results in Fig.~\ref{fig:diffP}(d) as the case of a reduced $g$ while fixing the other parameters. It is clearly seen that, with a reduced $N$ (thus, a reduced $g$), the exceptional coupling appears at a wider $w$. Reducing the $g$ increase the modal overlap, thus a wider $w$ (i.e., a higher confinement) is required to compensate for the increased modal overlap. Again, this is consistent with the parametric studies in Fig.~\ref{fig:parameter}(a).

\bibliographystyle{naturemag}
\nocite{*}